\documentclass[sigconf,authorversion,nonacm]{acmart}
\usepackage{microtype}
\usepackage{paralist}
\usepackage{enumitem}

\setlist[itemize]{leftmargin=10pt}
\setlist[enumerate]{leftmargin=13pt}

\begin{document}

\title{Teaching Software Engineering for AI-Enabled Systems}

\author{Christian K\"astner}
\orcid{0000-0002-4450-4572}
\affiliation{Carnegie Mellon University}
\author{Eunsuk Kang}
\affiliation{Carnegie Mellon University}

\begin{abstract}
\looseness=-1
Software engineers have significant expertise to offer when building
intelligent systems, drawing on decades of experience and methods
for building systems that are scalable, responsive and robust, even when built on unreliable components.
Systems with artificial-intelligence or machine-learning (ML) components
raise new challenges and require careful engineering.
We designed a new course to teach software-engineering skills
to students with a background in ML.
We specifically go beyond traditional ML courses that
teach modeling techniques under artificial conditions and focus,
in lecture and assignments, on realism with large and changing datasets,
robust and evolvable infrastructure, and purposeful requirements
engineering that considers ethics and fairness as well.
We describe the course and our infrastructure and share experience and all material from teaching the course for
the first time. 
\end{abstract}
\maketitle

\section{Introduction}

More and more modern software systems include  machine-learning (ML) models
for part of their functionality (e.g., recommendation engines in e-business
sites) or are even built around such models (e.g., mobile apps for instant
language translations). Artificial intelligence (AI), including the subfields
of ML and data analytics, are hot topics of interest to many of our students.
Yet, while AI courses abound, including more formal and more practical courses
and many online MOOC-style offerings and tutorials, we find that little
attention is paid to software-engineering aspects in building complete systems
that involve AI.

AI education typically focuses on algorithms and techniques
or on applying these techniques in artificial settings
(e.g., fixed datasets
and Jupyter notebooks), narrowly focused on optimizing
model accuracy.
However, for building real systems, many additional
challenges become important, for example: 
\emph{How to build robust AI pipelines and facilitate regular model updates?
How to deploy and update models in production?
How to evaluate data and model quality in production?
How to deal with mistakes that the model makes and manage associated risk?
How to trade off between various qualities, including learning cost, inference time, updatability, and interpretability?
%How to conduct experiments at scale?
How to design a system that scales to large amounts of data?
How to version models and data?}

\looseness=-1
Faced with an increasing interest on AI topics
from our students and little
offerings that address engineering concerns in 
AI courses, we decided to design a new course that would
address this niche: \emph{Software Engineering for AI-Enabled Systems.}
This paper reports on design considerations for a course 
that teaches software engineering
techniques for building systems with AI components
and experience from teaching the course for the first
time. It is meant to start a discussion and provides teaching materials
(including slides, exercises, and assignments)
as sharable artifacts released under a Creative Commons license at \url{https://github.com/ckaestne/seai/}.
Our technical simulation infrastructure
for the assignments is available on request.

\begin{figure}[tb]
	\centering
\includegraphics[width=\linewidth]{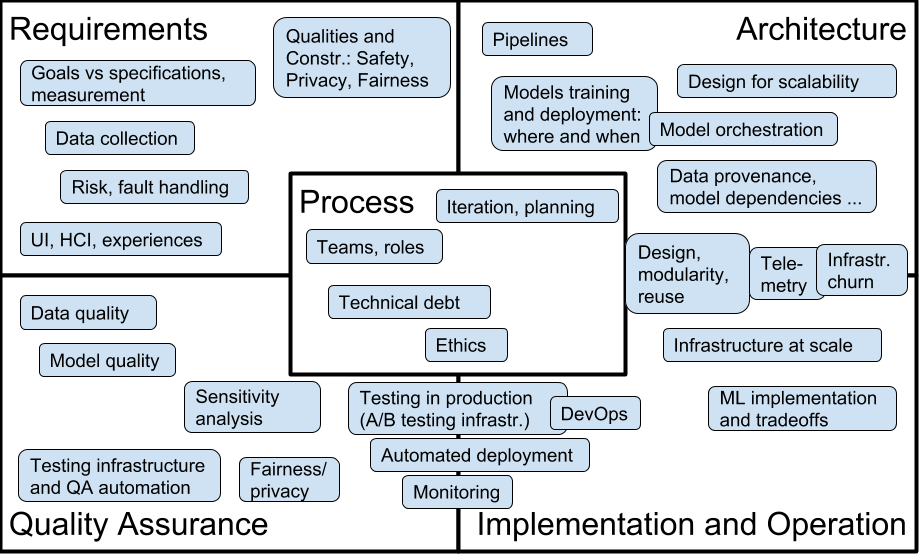}
\Description{Overview of course topics}
\caption{Overview of course topics}
	\label{fig:topics}
\end{figure}

\section{Related Work and Scope}
The software-engineering research community focuses 
primarily on how to use artificial intelligence techniques
to solve software engineering problems (AI4SE), for example,
for finding bugs~\cite[e.g.,][]{pradel2018deepbugs}, triaging bugs~\cite[e.g.,][]{tian2013drone}, or 
repairing bugs~\cite[e.g.,][]{LNFW:TSE12}.
For this line of research, there are also corresponding graduate courses taught by
software engineering researchers, for example, Breaux's \emph{Artificial Intelligence for Software Engineering} at Carnegie Mellon %~\cite{breaux-course-website}
or Siegmund's \emph{Search-Based Software Engineering} at the University of Weimar.
Our new course has the \emph{opposite
focus}: how software engineering techniques can be
used to build better systems with or around AI components (SE4AI).

Software engineering concerns for AI-enabled systems
arise in practice when deploying and operating actual systems.
Practitioners in large companies discuss their problems
and custom solutions in blog posts, talks, and position papers:
For example, Google teams discusses 
engineering challenges~\cite{Sculley2011-at},
technical debt~\cite{Sculley_undated-bb}, 
testing challenges~\cite{Breck2017-fj},
tracking data provenance~\cite{Halevy_undated-je}, and 
A/B testing~\cite{Tang_undated-lf} and
Uber discusses their learning platform in a series of blog posts~\cite{uber};
Microsoft studied the role of data analysts on their software teams~\cite{Kim2018-dn}
as well as challenges in cross-sectional teams~\cite{seml-seip19}.

On the academic side, we found only limited software-engineering research 
specific to AI-enabled systems.
\citet{Arpteg2018-jr} interviewed stakeholders from seven software
projects using deep learning to characterize challenges.
Some researchers focused on testing ML implementations~\cite{Srisakaokul_undated-re}
and ML models~\cite{Xie2011-ru,galhotra2017fairness}, with a recent interest in testing
and test coverage
of neural networks~\cite[e.g.,][]{ma2018deepgauge}.
In contrast, requirements, software architecture, and process seem rarely
discussed in the academic literature when it comes to AI-enabled systems.

While most books on AI focus on techniques and modeling,
\citet{mlsystems} and \citet{hulten} have written books that focus
more on engineering aspects. Smith's \emph{Machine Learning Systems}~\cite{mlsystems}
focuses on technical implementations and commits
on many specific design decisions. In contrast, Hulten's
\emph{Building Intelligent Systems}~\cite{hulten} actually covers
many design considerations fairly broadly and is the closest
coverage we found for the intended course. We assign chapters of Hulten's book
as required reading throughout the course.

\section{Course Design}

We design the course with a specific tension between data scientists and
software engineers in mind:
% \begin{itemize}
	% \item 
	\emph{Data scientists} often make great progress at building models with cutting edge techniques, but turning those models into products is challenging. For example, data scientists may work with unversioned notebooks on static data sets and focus on optimizing model accuracy while ignoring scalability, robustness, update latency, and operating cost.
	% \item 
	\emph{Software engineers} are trained to work with specifications and tend to focus on code, but may not be aware of the difficulties of working with data and unreliable models. They have a toolset for decision making, risk management, and quality assurance but it is not always obvious how to apply those to intelligent systems and their challenges.
% \end{itemize}
Our course adopts a \emph{software-engineering perspective} on building AI-enabled systems, focusing on what a software engineer can do to \emph{turn a machine learning idea into a scalable and reliable product}. %Rather than focusing on modeling and learning itself, i
It assumes a working relationship with a data scientist and focuses on issues of design, implementation, operation, and quality assurance.% and how those interact with the data scientist's modeling.

While there are distinct characteristics of AI components (especially
ML models), they
also relate to core topics in software engineering, for example:
\begin{itemize}
	\item ML components are used for problems for which 
	we cannot specify a solution,
	because the specification would be too complex or because it is unknown. 
	Instead of providing a specification, 
	one sets a goal and trains a model to figure out the answer.
	With this shift from deductive to inductive reasoning, 
	we give up expecting guarantees, but embrace best-effort solutions,
	fully accepting that answers can be wrong.
	However, while we often teach the value of clear and formal specifications,
	in practice, software engineers already routinely deal with underspecified 
	and unreliable components. We have developed techniques to build reliable and safe systems from unreliable or untrusted
	components -- those become essential when building AI-enabled
        systems.
      \item The
        environment plays a critical role in establishing requirements
        of an AI-enabled system.  Evaluating the long-term impact of a
        feedback loop or potential harm caused by a biased ML model,
        for example, involves identifying relevant stakeholders, their
        motivations, and interactions among them. Software engineers
        distinguish between the machine and the
        world, identify environmental
        assumptions, and evaluate quality in the context of
        the environment -- concepts that are also important for
        building AI-enabled systems.
	\item ML components can have non-local and non-monotonic effects.
	Local improvements may degrade other parts and their various
	qualities, while identifying the source of a problem can be challenging due to unclear
	provenance. In software engineering, failures in modularity and compositionality are well understood, e.g., with regard to feature interactions, calling for
	 careful design and system-level
	testing. Similarly, a robust 
	architecture and quality assurance regime (beyond model
        accuracy) plays an
	important role in AI-enabled systems.
	% , with many opportunities to be learned
	% from software architecture, continuous integration, and interaction and fuzz testing.
	\item Evaluating model quality on a training set is well established and increasingly
	attention is paid to bias and fairness, but models may perform differently
	in production, and data drift and adversaries may further degrade performance or
	compromise the system. Software engineers have developed many techniques to
	monitor systems, evaluate systems in production, and automate decisions,
	including A/B testing and continuous deployment -- again such techniques
	become essential when building AI-enabled systems.
	\item Many models become large and expensive to learn, to use, and to version.
	But again, we collected a vast body
	of knowledge on building and operating distributed and scalable systems,
	tracking revisions and variants, and manage configurations at scale --
	which designers of AI-enabled systems can benefit from.
\end{itemize}
Overall, traditional software engineering projects come in different sizes and complexity, 
in a spectrum from small and well-understood projects to large and complex projects.
We argue that AI-enabled systems have only few truly unique challenges, but 
nontrivial AI components tend to push such systems to the complex end of the spectrum of software projects,
often with important concerns about safety, risk, scalability, and robustness.
They share lots of challenges with other complex and large-scale software projects 
and can benefit from many corresponding techniques. 
We postulate that,
\emph{while developers of simple traditional systems may get away with poor practices,
most developers of AI-enabled systems will not} --
hence, educating developers of AI-enabled systems 
in state-of-the-art software engineering techniques is an important educational mission.

\subsection{Scope and Lectures}
To design the course, we settled on the following assumptions and scoping rules:
\begin{enumerate}
	% \item \emph{Claim:} Software engineers should have something to say about how to design systems with AI components, and it is different to some degree or requires a different focus from traditional software engineering.
	\item Explicit transfer of software engineering concepts to AI-enabled systems using software-engineering terminology, techniques and structure (e.g. test coverage, architecture views, fault trees).
	\item AI components largely considered as a black box with minimal discussion of internals, focusing on tradeoffs (e.g., decision trees vs.\ neural networks vs.\ symbolic AI); 
	little focus on steps in the data analytics process beyond necessary basics;
	not competing with technical AI courses. 
		% Specifically, only a basic understanding needed to understand engineering tradeoffs among different techniques (latency, cost, data requirements, explainability, ...), but not competing with technical AI courses; little focus on steps in the data analytics process beyond necessary basics (e.g., how to identify features and normalize data for a specific learning technique
		% is not covered).
	\item Practice design decisions and analysis around concrete scenarios; hands-on experience with implementing the plumbing (e.g, monitoring, automated deployment, containers).
\end{enumerate}

We identified topics around all stages of the software engineering lifecycle, as
summarized in Figure~\ref{fig:topics}. Broadly, the lectures cover:
\begin{itemize}
	\item \emph{Requirements:} Understanding system goals; the lack of specifications for AI-components; identifying and measuring qualities of interest (beyond model accuracy), setting expectations for safety, security, and fairness; hazard analysis and fault tress; planning how to deal with mistakes.
	\item \emph{Architecture:} Considering tradeoffs among quality attributes (e.g., learning time, inference latency, model size, updatability, interpretability); planning where and how to deploy an AI-component; planning telemetry; data provenance; model orchestration, service-oriented architectures.
	\item \emph{Implementation and operation:} Designing scalable distributed systems for data and computation; infrastructure for experimentation, A/B testing, canary releases, and continuous delivery; provenance and configuration management; system monitoring.
	\item \emph{Quality assurance:} Measuring model quality offline and in production;  assuring data quality; testing the entire ML pipeline; safety, security, and fairness analysis.
	\item \emph{Process:} Iteration and planning; working with interdisciplinary teams;  technical debt; ethical decision making.
\end{itemize}

\subsection{Assignments}
A key design goal of this course is to provide hands-on software engineering
experience and to move away from merely building and evaluating models on
static datasets (e.g., from Kaggle.org). 
That is, we need to create a setting in
which many production concerns can be addressed, such as scaling
for large amounts of data. % and requiring high availability despite updates.
In addition, to allow assignments around A/B testing,
canary releases, or detecting feedback loops, we need an environment
in which predictions that students make with their models have an
actual effect on the environment.
To that end, we decided to design core technical assignments around
a simulation infrastructure.

\emph{Simulation Infrastructure.} The key idea is to build a
simulation infrastructure with a secret ground-truth model of the
world to simulate the behavior of environmental entities (e.g.,
users on a video streaming app). Students do not
have access to this model but can partially observe the simulated
world through shared events, and thus learn their own models to
represent the world.  The simulation then reacts to predictions that
student teams provide through an API.  With this design, we have
control of all produced data and can scale the simulation to create
suitable amounts of data, and, more importantly, simulate feedback
loops by having the simulated world repeatedly react to the student
predictions over time.

We have built such simulation for the scenario of a movie-streaming service
(think \emph{Netflix}).
We simulate several thousand users picking, watching, and rating movies,
creating a stream of watching and rating events.
We use an existing large dataset of movie ratings~\cite{harper2016movielens} 
to build our ground-truth
model of movie preferences for each user (hidden from students).
We import movie data from the original dataset and
use models to create artificial but representative data about all our users
(e.g., age, gender, occupation, activity level).
In our simulation, at the beginning of each day, we decide when and which users are going
to watch a movie and enqueue them in a timed event queue.
At the start time, the system will request movies recommendations for that user from
the student API and use the ground-truth model to select a movie among
the recommended movies and a random sample of other movies, picking the movie
that (after adding some noise and a small bonus for recommended movies) 
has the highest predicted rating. We then queue a sequence of public events that represent the user watching the movie, followed by an (optional)
rating event in which the simulated user reveals their real rating for the movie
(again derived from the ground-truth model, random noise, and a small bonus to favor
watched movies).
The ground-truth model is occasionally updated based on recent movie watching history.

The students interact with our simulator through (i) a stream of events in
an \emph{Apache Kafka} server to which they can subscribe (movie watch
event, rating events, and logs about received recommendations),
and (ii) an \emph{REST API} to query information about movies
and users. In addition, we specify the interface for the
recommendation service that they need to implement (REST API).

The system is designed such that students can evaluate their models in production,
e.g., how frequently do users watch recommended movies, do they finish them, and how do they rate them?
In addition, it can exhibit feedback loops, e.g., students constantly recommending horror 
movies will likely lead to many users developing a taste for horror movies.
Furthermore, we can simply hardcode data drift and schema changes or occasionally provide corrupt data to challenge
the robustness of the student's infrastructure.
The ground-truth model can use hidden information and specifically encode biases--for example, wrong age data combined with adolescent users strongly preferring R-rated movies.
This environment is also flexible enough to illustrate how different goals
(e.g., maximizing profit rather than
maximizing top ratings) can lead to very different outcomes and unintended side effects.

\emph{Assignments.} 
For the first year, we created a series of five group assignments that
build on the common movie streaming scenario:
\begin{enumerate}
	\item \emph{Modeling basics and offline evaluations:} Collect data from
	the system (Kafka stream and API) and build and evaluate a model
	(usually collaborative filtering)
	to get familiar with the infrastructure, practice basic 
	ML skills, and onboard all team members.
	\item \emph{Tradeoff analysis:} Focus on measurement of various qualities (beyond model accuracy) by trying and 
	comparing different modeling techniques and empirically discovering their tradeoffs.
	\item \emph{Infrastructure deployment and testing:}
	Migrate the solution from a Jupyter Notebook to a robust and scalable learning
	and inference infrastructure, build checks for
	data quality and test the entire infrastructure,
	assess model quality in production, and 
	deploy the model as a REST API using Docker containers.
	\item \emph{Model updates:} 
	Fully automate model updates (continuous deployment, including automated canary releases),
	deploy updates without downtime,
	and perform experiments with A/B tests in production using
	a self-developed infrastructure.
	\item \emph{Feedback loops:} Analyze the system for potential
	feedback loops and attack scenarios, design interventions, and
	continuously monitor system performance.
\end{enumerate}

In addition, we created 5 smaller individual assignments focusing on
(1)~identifying engineering concerns in a report about an AI-enabled
system, (2)~identifying safety requirements and using fault trees to
analyze a self-driving car accident, (3)~modeling and discussing
architecture tradeoffs regarding when and where to deploy and update
models for smart dashboard cameras, (4)~discussing and measuring
fairness concerns in a credit rating dataset, and (5)~analyzing
security vulnerabilities of an AI-based system using threat modeling.

\section{Experience}

We taught this class for the first time in the Fall 2019 semester to a small group of 12 graduate students.
From this experience, we can derive a number of recommendations for future semesters and others
wishing to teach such a class.

\emph{Focus and prerequisites:} We initially planned the class to require both basic knowledge
in ML and some software-engineering experience. This dual requirement limits the target audience of the class
and students actually taking the course often have gaps in either area so that we need to repeat many concepts. 
We believe that the better solution would be to teach separate sections for students 
from either background -- for ML students, we'd introduce software engineering
concepts from scratch (e.g., continuous integration, versioning, software architecture) not making assumptions
about prior experience; for software-engineering students we could teach a class that
begins with a pragmatic introduction to ML pipelines and model quality measures.
For our Master's program in software engineering, the content could
ideally be broken up into smaller pieces that could
be integrated as modules in our existing software engineering classes on requirements, software architecture, and quality assurance.

\looseness=-1
\emph{Simulator engineering:} Building and running the simulator required nontrivial
engineering effort and there were many features that we could not implement in the first offering due to time
constraints. For example, the watch behavior in our simulator is not particularly realistic (e.g., we do not have
power users who binge multiple movies, we do not have a model for stopping, restarting, or rewinding movies,
we do not explicitly model  demographics or locations of users).
We also found that the used scale with 160,000 users producing an average of 70 events per second
and 1 recommendation request per second is way too low to challenge students to seriously consider 
operating cost and performance, resulting sometimes in rather superficial engineering tradeoffs with obvious answers.
The ML task of recommending movies is also not computationally challenging enough
to offer interesting tradeoff discussions among different learning techniques.
In future offerings, it may be worth scaling the simulator to many more users and to
explore additional learning tasks that involve pictures, audio, or video.

\emph{Practical grounding:}
In addition to the movie recommendation scenario used for homework, we use different 
scenarios in almost every single lecture to discuss the breadth of different problems
and the importance of making system-specific design and tradeoff decisions.
At the same time, it may be worth exploring a few scenarios or even concrete implementations in more depth. Another way to ground the course more in practice is
to invite more guest speakers that use AI in their systems.

\emph{Tooling:} As AI is an active, rapidly evolving field, 
we struggled with finding standard techniques or mature tools for
emerging topics such as fairness and explainability. In addition, the
shift of focus from code to data brings about an increasing demand for
software-engineering tools for data-intensive tasks, such as
versioning of large datasets, data cleaning, and model quality
evaluation. Although a few tools exist for these tasks (e.g., the
What-If tool by Google), most are still experimental and challenging 
to use for teaching. We believe that the software
engineering community -- with its expertise in tool development and
experimentation -- has a plenty of opportunities to contribute to the growing needs
of AI-enabled system developers and educators by developing a set of
mature tools and benchmarks.

\section{Conclusion}
Systems with an AI component are challenging to build and to maintain.
We designed a course to teach software engineering to students interested
in AI to foster broader thinking beyond a narrow focus on static
datasets and model quality.
We shared our experience and make all course material available under creative commons license.

\paragraph{Acknowledgments.}
We would like to thank all colleagues and practitioners who encouraged and
supported us in creating this course and provided feedback on its content, including
Travis Breaux,
Owen Cheng,
Fei Fang,
David Garlan,
Michael Hilton,
Geoff Hulten,
Pooyan Jamshidi,
and
Norbert Siegmund.
We also greatly appreciate the help of 
Yasasvi Hari,
Dong Won Lee, 
Chu-Pan Wong,
and
Zhendong Yuan
in designing and implementing the course infrastructure.
Also thanks for the students in the first iteration of
the course for sticking with us and providing feedback.

\bibliographystyle{ACM-Reference-Format}
\bibliography{dblp2_short}

\end{document}